\documentclass[sigconf]{acmart}

\renewcommand\footnotetextcopyrightpermission[1]{}
\usepackage{amsmath,amsfonts}
\usepackage{graphicx}
\usepackage{textcomp}
\def\BibTeX{{\rm B\kern-.05em{\sc i\kern-.025em b}\kern-.08em
    T\kern-.1667em\lower.7ex\hbox{E}\kern-.125emX}}

\usepackage{lipsum}

\usepackage[shortlabels]{enumitem}
\setlist[itemize]{align=parleft,left=0pt..1em}
\setlist[enumerate]{align=parleft,left=0pt..1.5em}

\definecolor{turquoise}{rgb}{0.1, 0.75, 0.6}

\newcommand{\red}[1]{\textcolor{red}{#1}}

\newcommand{\blue}[1]{\textcolor{blue}{#1}}
\newcommand{\magenta}[1]{\textcolor{magenta}{#1}}

\newcommand{\orange}[1]{\textcolor{orange}{#1}}

\newcommand{\xian}[1]{{\textcolor{violet}{#1}}}
\newcommand{\revise}[1]{\textcolor{cyan}{#1}}

\usepackage{pifont}
\newcommand{\mycomment}[2]{[\ding{46}~#1:~#2]}
\newcommand{\sinan}[1]{\blue{\mycomment{sinan}{#1}}}
\newcommand{\yepang}[1]{\red{\mycomment{yepang}{#1}}}
\newcommand{\xiancm}[1]{\magenta{\mycomment{xian}{#1}}}
\newcommand{\ying}[1]{\orange{\mycomment{Ying}{#1}}}

\newcommand*{\mytt}{\fontfamily{lmtt}\selectfont}

\usepackage{listings}
\definecolor{javared}{rgb}{0.6,0,0}
\definecolor{javagreen}{rgb}{0.25,0.5,0.35}
\definecolor{javapurple}{rgb}{0.5,0,0.35}
\definecolor{javadocblue}{rgb}{0.25,0.35,0.75}
\definecolor{commentgrey}{rgb}{0.75,0.75,0.75}
\definecolor{diffred}{rgb}{0.8,0,0}
\definecolor{diffgreen}{rgb}{0,0.7,0}
\lstdefinelanguage{diff}{
  basicstyle=\linespread{0.8}\mytt\small,
  morecomment=[l][\color{commentgrey}]{//},
  morecomment=[f][\color{diffred}]{-},       
  morecomment=[f][\color{diffgreen}]{+},     
}

\usepackage{csquotes}
\renewcommand{\mkbegdispquote}[2]{\itshape}

\usepackage{subcaption}

\usepackage{multirow}
\usepackage[flushleft]{threeparttable}

\usepackage[
ruled,
noend,
linesnumbered,
]{algorithm2e}
\makeatletter
\newcommand{\algorithmfootnote}[2][\footnotesize]{%
  \let\old@algocf@finish\@algocf@finish
  \def\@algocf@finish{\old@algocf@finish
    \leavevmode\rlap{\begin{minipage}{\linewidth}
    #1#2
    \end{minipage}}%
  }%
}
\makeatother

\usepackage{stmaryrd}

\usepackage{comment}

\usepackage{hyperref}

\newcommand{\tool}{\textsc{Aper}~}
\newcommand{\tooln}{\textsc{Aper}}
\newcommand{\bench}{\textsc{ARPfix}~}
\newcommand{\benchn}{\textsc{ARPfix}}

\newcommand*{\ie}{i.e.}
\newcommand*{\eg}{e.g.}

\newcommand*{\etal}{et al.}

\newcommand*{\placeholder}{\textcolor{red}{xxx~}}

\begin{document}

\title{\tooln: Evolution-Aware Runtime Permission Misuse Detection for Android Apps}
\thanks{This work has been accepted to ICSE 2022.}

\author{Sinan Wang}
\affiliation{
  \institution{Southern University of Science and Technology}
  \city{Shenzhen}
  \country{China}
}
\email{wangsn@mail.sustech.edu.cn}

\author{Yibo Wang}
\affiliation{
  \institution{Northeastern University}
  \city{Shenyang}
  \country{China}
}
\email{yibowangcz@outlook.com}

\author{Xian Zhan}
\affiliation{
  \institution{The Hong Kong Polytechnic University}
  \city{Hong Kong}
  \country{China}
}
\email{chichoxian@gmail.com}

\author{Ying Wang}
\affiliation{
  \institution{Northeastern University}
  \city{Shenyang}
  \country{China}
}
\email{wangying@swc.neu.edu.cn}

\author{Yepang Liu}
\affiliation{
  \institution{Southern University of Science and Technology}
  \city{Shenzhen}
  \country{China}
}
\email{liuyp1@sustech.edu.cn}

\author{Xiapu Luo}
\affiliation{
  \institution{The Hong Kong Polytechnic University}
  \city{Hong Kong}
  \country{China}
}
\email{csxluo@comp.polyu.edu.hk}

\author{Shing-Chi Cheung}
\affiliation{
  \institution{The Hong Kong University of Science and Technology, and Guangzhou HKUST Fok Ying Tung Research Institute}
  \city{Hong Kong}
  \country{China}
}
\email{scc@cse.ust.hk}

\begin{abstract}
The Android platform introduces the runtime permission model in version 6.0.
The new model greatly improves data privacy and user experience, but brings new challenges for app developers.
First, it allows users to freely revoke granted permissions. Hence, developers cannot assume that the permissions granted to an app would keep being granted. Instead, they should make their apps carefully check the permission status before invoking dangerous APIs.
Second, the permission specification keeps evolving, bringing new types of compatibility issues into the ecosystem.
To understand the impact of the challenges, we conducted an empirical study on 13,352 popular Google Play apps.
We found that 86.0\% apps used dangerous APIs asynchronously after permission management and 61.2\% apps used evolving dangerous APIs.
If an app does not properly handle permission revocations or platform differences, unexpected runtime issues may happen and even cause app crashes.
We call such Android Runtime Permission issues as ARP bugs.
Unfortunately, existing runtime permission issue detection tools cannot effectively deal with the ARP bugs induced by asynchronous permission management and permission specification evolution.
To fill the gap, we designed a static analyzer, \tooln, that performs reaching definition and dominator analysis on Android apps to detect the two types of ARP bugs.
To compare \tool with existing tools, we built a benchmark, \benchn, from 60 real ARP bugs.
Our experiment results show that \tool significantly outperforms two academic tools, \textsc{ARPDroid} and \textsc{RevDroid},
and an industrial tool, \textsc{Lint}, on \benchn, with an average improvement of 46.3\% on $F_1$-score.
In addition, \tool successfully found 34 ARP bugs in 214 open-source Android apps, most of which can result in abnormal app behaviors (such as app crashes) according to our manual validation.
We reported these bugs to the app developers.
So far, 17 bugs have been confirmed and seven have been fixed.
\end{abstract}


\maketitle
\pagestyle{plain}

\section{Introduction}


The permission mechanism on the Android platform serves as an essential guard to users' data privacy.
In order to access sensitive data or critical system functions, an Android app should obtain corresponding permissions from the user before invoking related permission-protected APIs.
Prior to Android~6.0 (i.e., API level 22 or earlier, which we call ``\textit{legacy platforms}''), the permission granting process happens when an app is installed \cite{wang2015reevaluating}.
After the app is granted with its required permissions and gets installed, users can no longer revoke the permissions\footnote{A few customized legacy platforms allow users to revoke permissions after installation~\cite{WEB:stackoverflow/below23/xiaomi}. They are out of the scope of this work.}.
Such a static permission model was considered to be vulnerable, as the installation-time permission warnings cannot effectively help users make their security decisions~\cite{felt2012android}.
To address this weakness, in Android~6.0 (API level 23), the \textit{runtime permission model} was introduced.
The new model brought several changes: 1) \textit{dangerous permissions} are requested during an app's execution; 2)
users can either grant or deny such requests; 3) users can also revoke granted permissions from the system settings at any time.
These changes greatly improve data privacy and users' experience with Android apps~\cite{andriotis2016permissions}. However, this Android runtime permission (ARP) mechanism also brings technical challenges to developers:


\begin{itemize}
	\item As users can freely revoke granted permissions starting from Android 6.0, developers cannot guarantee that the required permissions are always held by their apps. To avoid app crashes, they need to properly insert permission check and request statements to ensure every dangerous API used by their apps is never invoked without the required permissions~\cite{WEB:android/request-permission}.
	This is a non-trivial task, even for experienced developers.
	For example, in issue 2110~\cite{WEB:github/k-9/2110} of \textit{K-9 Mail}, a famous email client for Android, the developers had discussed how to support runtime permissions for almost two years, until they ``functionally covered the user experience of requesting (runtime) permissions''.
	
	\item Apart from moving the permission granting process to runtime, the Android platform has also changed the behavior of permission groups~\cite{calciati2020automatically}, altered many APIs' corresponding permissions~\cite{WEB:github/WiFiAnalyzer/202}, and supported one-time permissions~\cite{WEB:android/11-updates}.
	Without tracking and properly handling such changes, app developers cannot easily implement permission-protected functionalities~\cite{wang2021runtime}.
\end{itemize}

Previous work mostly studied how to adapt apps targeting legacy platforms to the new platforms~\cite{bogdanas2017dperm,dilhara2018automated,gasparis2018droid}, or detect ARP issues from the security perspective~\cite{sadeghi2018temporal}.
The proposed tools cannot effectively detect ARP-induced functional bugs \cite{fang2016revdroid,sadeghi2017patdroid} for two reasons:
\begin{itemize}
\item Permission-protected APIs and their permission specification keep evolving, which is overlooked by almost all existing tools, causing many ARP-related compatibility issues undiscovered.
\item Most existing tools, like \textsc{ARPDroid}~\cite{dilhara2018automated} and \textsc{RevDroid}~\cite{fang2016revdroid}, leverage FlowDroid's dummy main classes \cite{arzt2014flowdroid} to model the implicit control flows within each component of an Android app. 
However, as developers may manage runtime permissions asynchronously across app components, performing such an intra-component analysis would produce false alarms.
\end{itemize}

To further understand the limitations of the existing tools and the practices of ARP management, we conducted an empirical study by analyzing the source code of the Android platform and 13,352 popular Android apps.
The results show that a large number of apps may be affected by the active changes of dangerous permissions and dangerous APIs, and asynchronous permission management is common in real-world apps.

Driven by our empirical findings, we designed and implemented \tooln, an \underline{\bf A}ndroid runtime
\underline{\bf P}ermission misus\underline{\bf E} bug detecto\underline{\bf R}.
\tool is an evolution- and asynchrony-aware ARP bug detector. It performs both intra- and inter-component static analyses to find missing permission and runtime version checks.
To evaluate \tooln, we built a benchmark from 60 real ARP bugs.
The experiment results on the benchmark show that \tool outperforms existing tools by an average improvement of 46.3\% on \textit{$F_1$-score}.
To further evaluate the usefulness of \tooln, we applied it to analyze 214 open-source apps.
It successfully found 34 real ARP bugs.
We submitted bug reports to the app developers. At the time of paper acceptance, 17 submitted bugs have been confirmed by the developers and seven have already been fixed.
In summary, we make the following contributions:

\begin{itemize}

\item To the best of our knowledge, we conducted the first \textbf{empirical study} on the evolution of ARP specification and real-world developers' practices in runtime permission management.

\item We proposed a \textbf{static analyzer}, \tooln, for detecting ARP bugs in Android apps, with a special focus on evolution-induced issues.

\item We performed an \textbf{evaluation} of \tooln, including both control experiments on our prepared benchmark and an in-the-wild study on real-world open-source apps. The results show that \tool can significantly outperform existing tools.

\item We provided a \textbf{reproduction package} for future research at {\url{https://aper-project.github.io/}}, which includes:
1) \tooln's source code,
2) the benchmark of 60 ARP bugs and their patches,
and
3) 34 real ARP bugs detected by \tooln, along with our issue reproducing videos, detailed descriptions, and developers' feedback.

\end{itemize}

\section{Preliminaries}
\label{sec:preliminaries}

\subsection{Permission Specification}
\label{sec:permission_spec}

Android framework provides APIs to perform sensitive operations, such as accessing contacts or using the camera.
These APIs are protected by corresponding \textit{dangerous permissions}, and we call them \textit{dangerous APIs}.
For instance, to open the camera, an app should request the user to grant the {\mytt CAMERA} permission before invoking the dangerous API {\mytt CameraManager.openCamera()}.
Unlike dangerous permissions, \textit{normal permissions} (\eg, network, vibration) can be automatically granted when the app is installed \cite{sadeghi2018temporal}.
Generally, Android permissions have four protection levels \cite{zhauniarovich2016small}: \textit{normal}~$\prec$~\textit{dangerous}~$\prec$~\textit{signature}~$\prec$ ~\textit{signatureOrSystem}, where $a \prec b$ means that the permission with a protection level $b$ has a higher risk than that of $a$, and thus granting them should follow different procedures. 

\begin{figure}[t]
\begin{lstlisting}[language=java]
    @RequiresPermission(anyOf={ACCESS_COARSE_LOCATION,
                                  ACCESS_FINE_LOCATION})
    public void requestLocationUpdates(...) { ... }
\end{lstlisting}
    \vspace{-1em}
    \caption{Permission specification of a dangerous API}
    \label{fig:requestLocationUpdates}
    \vspace{-2ex}
\end{figure}

In Android framework, starting from 6.0, a permission-protected API can use the {\mytt @RequiresPermission} annotation to specify its required permissions.
Figure \ref{fig:requestLocationUpdates} shows the API {\mytt requestLocation Updates()}, and its annotation-based permission specification.
According to this annotation, any of the {\mytt ACCESS\_COARSE\_LOCATION} or the {\mytt ACCESS\_FINE\_LOCATION} permission should be granted before invoking this dangerous API.
Similarly, if all listed permissions are required by the API, they will be specified by the element {\mytt allOf}, instead of {\mytt anyOf}.
Such annotations can help developers and static checkers (\eg, \textsc{Lint} \cite{wei2017oasis}) to determine whether an API is protected by any permissions.
Besides the annotation, the permissions can also be implicitly specified using the {\mytt \{@link android.Manifest. permission\#...\}} tag in an API's Javadoc.

\subsection{Runtime Permission Management}
\label{ssec:management}

\begin{figure}[t]
    \resizebox{\columnwidth}{!}{ 
\begin{lstlisting}[language=java]
if(checkSelfPermission("ACCESS_FINE_LOCATION") == GRANTED)
  doLocationingActions(...); // calls dangerous API
else
  requestPermissions(new String[]{"ACCESS_FINE_LOCATION"});
\end{lstlisting}}
    \vspace{-1em}
    \caption{Permission check and request example (simplified)}
    \label{fig:requestPermissions}
    \vspace{-2ex}
\end{figure}

To acquire a permission, an Android app should declare it in the manifest file using the {\mytt <uses-permission>} XML element.
This, however, does not guarantee that the permission can always be granted after the app is installed.
The legacy platforms provide APIs for apps to check their permission statuses at runtime.
Under the runtime permission model, users are allowed to dynamically grant and revoke dangerous permissions.
Thus, the new platforms (API level 23 and above) provide additional APIs to handle users' runtime behaviors.
Figure \ref{fig:requestPermissions} shows an example from the official documentation \cite{WEB:android/request-permission}, which involves permission check and request.
Generally, the permission management APIs fall into four categories:

\textbf{(a) \textit{Checking permission status}}:
Before invoking a permission-protected API, an app will check whether it has the required permissions by invoking the \textit{CHECK} APIs (\eg, {\mytt ContextCompat.check SelfPermission()}).
These APIs typically accept a permission string and return whether the permission is granted or not.

\textbf{(b) \textit{Requesting for dangerous permissions}}:
To request dangerous permissions at runtime, an app will call \textit{REQUEST} APIs like {\mytt ActivityCompat.requestPermissions()}.
This will trigger a pop-up dialog that prompts the user to either grant/deny the permission request, or block subsequent permission requests.
Unlike \textit{CHECK} APIs, the \textit{REQUEST} APIs accept an array of permission strings to spawn multiple permission request dialogs.

\textbf{(c) \textit{Handling user response}}:
The \textit{HANDLE} API, {\mytt onRequest PermissionsResult(int,String[],int[])}, is an empty callback defined in the base GUI classes.
It is invoked by the system after the user reacts to the permission request, and its parameters store the user's granting results.
Developers can override this callback to check the user's decisions and take actions accordingly.

\textbf{(d) \textit{Explaining permission usage}}:
The \textit{EXPLAIN} API, {\mytt should ShowRequestPermissionRationale()}, returns a boolean value of whether the user has denied the permission request and selected the ``Never ask again'' option.
Developers may use it to check whether the permission requests are blocked, and explain to the user why the requested permission is essential, accordingly.

\section{Empirical Study}
\label{sec:empirical}

To understand the practices of runtime permission management and the limitations of existing work, we conducted an empirical study to investigate the following three research questions:

\begin{itemize}
\item \textbf{RQ1 (Evolution of Permission Specification)}: How do the permission-protected APIs and their permission specification evolve in the Android platform?
\item \textbf{RQ2 (Impact of Evolution)}: How many Android apps may be affected by the evolution of API-permission specification?
\item \textbf{RQ3 (Permission Management Practices)}: How many Android apps implement asynchronous permission managements?
\end{itemize}

In the following, we present our data collecting procedures, analysis methods, and results.

\subsection{Dataset Construction}
\label{ssec:dataset}

\subsubsection{Collecting API-Permission Mappings}

For RQ1, we collected the API-permission mappings from the Android framework~\cite{bogdanas2017dperm}.
For each API level, we traversed all Java files to find those API methods that have {\mytt @RequiresPermission} annotations or use {\mytt @link} tags to specify their permission requirements (\S~\ref{sec:preliminaries}).
The declared permissions are extracted from the framework's manifest file.
We did not adopt the existing mappings \cite{stowaway2011CCS,au2012pscout,backes2016demystifying,bogdanas2017dperm,aafer2018precise}, because their released datasets are outdated, or the mapping extraction tools are unavailable or cannot be easily applied to analyze the new Android versions.
More importantly, as discussed in a recent work~\cite{dawoud2021bringing}, these mappings are neither precise nor complete.
Since our study needs the latest and precise API-permission mappings, we decided to extract mappings from the source code and docs.

\subsubsection{Collecting Android Apps}

RQ2 and RQ3 investigate the permission management practices in real-world apps.
To this end, we crawled Google Play \cite{WEB:google-play} apps from \emph{Androzoo} \cite{allix2016androzoo} according to two criteria:
1) ranking at top-500 in each of the 32 categories indexed by AppBrain~\cite{WEB:appbrain/google-play} (\ie, the apps should be popular);
2) containing call sites of both dangerous API(s) and permission management API(s). 
As we implemented our analyzer on top of Soot \cite{lam2011soot}, a well-maintained Java program analysis framework, we excluded all apps in the game category because they are mostly built using game engines that are not developed in Java \cite{nagappan2016future}. 

\begin{figure}[t!]
    \includegraphics[width=\columnwidth]{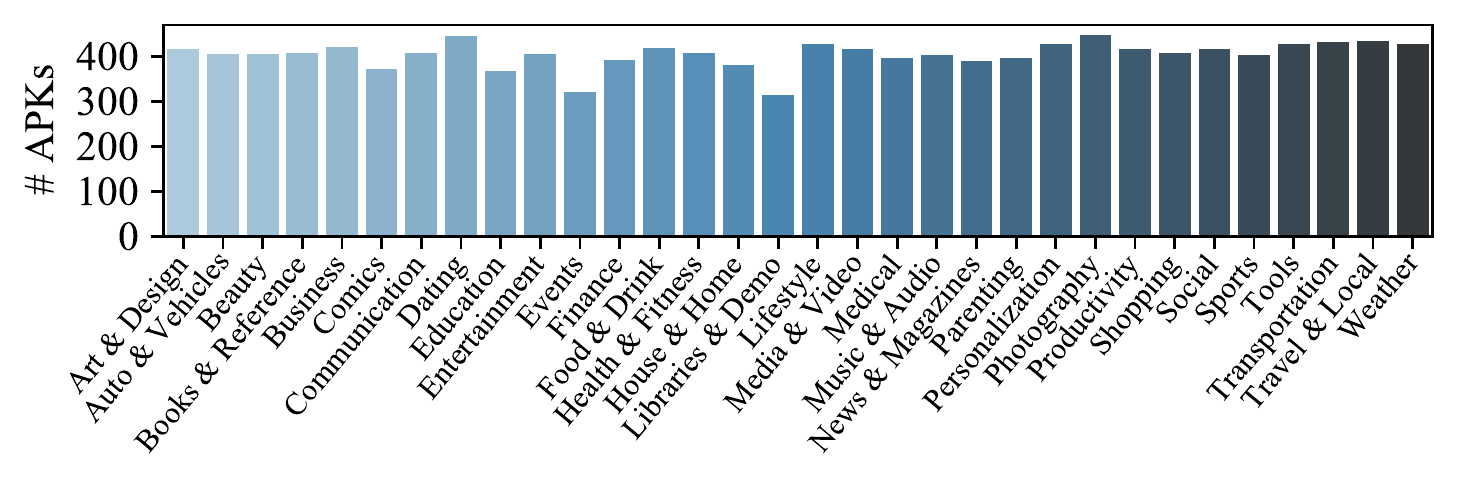}
    \vspace{-2.5em}
    \caption{Category distribution of the Google Play apps}
    \label{fig:gplay-cat}
    \vspace{-1em}
\end{figure}

Finally, we collected 13,352 popular apps that involve dangerous permission usages.
The category distribution of these apps is shown in Figure \ref{fig:gplay-cat}.
In general, our collected apps are evenly distributed among different categories, thus studying them can reveal developers' practices of ARP management without bias toward certain app categories.

It is worth noting that we only focus on \textit{dangerous APIs} in RQ2 and RQ3.
The reason is that normal and signature permissions do not require runtime management since they are granted at the time of installation.
An app can use any API protected by such permissions without users' explicit approvals.
In this case, the safety of an API call can be guaranteed by simply checking whether the required permissions are declared in the manifest file of the app.

\subsection{RQ1: Evolution of Permission Specification}
\label{ssec:rq1}

\begin{table*}[t]
\caption{Distribution and evolution of permission specification}
\label{tab:rq1}
\vspace{-1em}
\resizebox{\textwidth}{!}{
\begin{threeparttable}

\begin{tabular}{lrrrrrrrrrrrrr}
\toprule

\multirow{2}{*}[-0.3em]{\parbox{5.5em}{\centering\Large{API level}\linebreak\footnotesize{(version number)}}} & \multicolumn{3}{c}{\#Permissions} & \multicolumn{4}{c}{\#APIs protected by permissions} & \multicolumn{3}{c}{\#Evolved APIs} & \multicolumn{3}{c}{Change of permissions} \\

\cmidrule(lr){2-4}
\cmidrule(lr){5-8}
\cmidrule(lr){9-11}
\cmidrule(l){12-14}

&  normal &  dangerous &  signature &  total &  normal &  dangerous &  signature & added & deleted & changed$^\dagger$ & restricted & relaxed & same-level \\       

\midrule

23\ \ (6.0) &      55 &         25 &        235 &      452 &             260 &                 71 &                127 &         - &           - &           - &              - &           - & -\\
24\ \ (7.0) &      55 &         25 &        270 &      540 &             269 &                108 &                172 &        98 &          10 &           0 &              0 &           0 & 0\\
25\ \ (7.1) &      55 &         25 &        271 &      547 &             269 &                110 &                177 &        10 &           3 &           0 &              0 &           0 & 0\\
26\ \ (8.0) &      57 &         27 &        303 &      822 &             303 &                146 &                400 &       290 &          15 &           6 &              1 &           1 & 4\\
27\ \ (8.1) &      57 &         27 &        323 &      889 &             309 &                152 &                446 &        75 &           8 &           5 &              3 &           1 & 1\\
28\ \ (9)   &      60 &         28 &        369 &     1,094 &             345 &                162 &                620 &       230 &          25 &          12 &              3 &           2 & 7\\
29\ \ (10)  &      63 &         31 &        441 &     1,523 &             359 &                195 &               1,011 &       467 &          38 &          31 &             12 &           3 & 16\\
30\ \ (11)  &      63 &         31 &        497 &     2,036 &             364 &                218 &               1,504 &       580 &          67 &          86 &             24 &          10 & 52\\

\bottomrule
\end{tabular}

\begin{tablenotes}
    \item $^\dagger$ We say a permission-protected API is \textit{changed} if its required permissions are changed.
\end{tablenotes}

\end{threeparttable}

}
\end{table*}

\subsubsection{Analysis Method}




We studied permissions and permission-protected APIs, by investigating their distributions and how they have changed across Android versions from 6.0 to 11.
We grouped the APIs according to their required permissions' protection levels.
We also merged \textit{signature} and \textit{signatureOrSystem} permissions (\S~\ref{sec:permission_spec}) into one protection level, because they are restrictively used by specific apps and cannot be accessed by all developers \cite{zhauniarovich2016small}.

\subsubsection{Results}

Table \ref{tab:rq1} summarizes the evolution of permission specification from Android 6.0 to 11. 
Columns 2-4 show that the number of permissions generally increases as the Android platform evolves, regardless of their protection levels.
Signature permissions grow the fastest, which is in line with previous work~\cite{zhauniarovich2016small}.
Columns 5-8 show how many API methods are protected by permissions of specific protection levels.
Note that the mapping between API methods and permissions is not bijection. It is possible that an API requires multiple permissions \cite{au2012pscout}, or a permission is bound to a constant field rather than any methods.
Columns 9-11 report the evolution of permission-protected APIs in consecutive versions, in terms of their addition, deletion, and modification.
As Android platform evolves, more permission-protected APIs are added.
Meanwhile, the numbers of deleted/changed APIs are also increasing.
The permission changes of APIs are shown in the last three columns.
We say a changed API becomes \textit{restricted} if it requires a permission with a higher protection level in new versions, and \textit{relaxed} on the contrary.
Besides, an API can also switch to another permission with the same protection level (the last column).
As shown in columns 12-14, the changed APIs tend to restrict their permissions in new versions, or change to other permissions with the same protection level.
Only a few require permissions with lower protection levels.

From the results, we can see that {\bf dangerous permissions and APIs constantly evolve along with the Android platform}.
For example, 29 out of the 580 new APIs require dangerous permissions in Android 11, and seven dangerous APIs are deleted from Android 10.
For those APIs that change their required permissions, 20 are related to dangerous permissions (becoming either restricted or relaxed).
As dangerous APIs are commonly used to implement sensitive features, their usages and compatibilities should be carefully checked to avoid unexpected runtime behaviors.

\subsection{RQ2: Impact of Evolution}
\label{ssec:rq2}

\subsubsection{Analysis Method}
\label{sssec:method}

To study RQ2, we first analyzed all dangerous APIs since API level 23. In total, there were 246 dangerous APIs with different method signatures (up to API level 30). 
We found that 188 of them have undergone changes, which account for a large proportion.
When using such \textit{evolving dangerous APIs}, developers should carefully examine the running device versions
and deal with the permission changes to avoid unexpected program behaviors (\eg, crash) \cite{dilhara2018automated}.
RQ2 aims to quantify the potential impact of the evolving dangerous APIs on real-world apps.

Given an Android app, for each call site of the evolving dangerous API, we extracted all possible calling contexts~\cite{sumner2011precise}, which are paths in the call graph (CG) starting from an entry method, including lifecycle callbacks (\eg, {\mytt onCreate()}), event handlers (\eg, {\mytt onClick()}), threads' {\mytt Runnable.run()}, etc., and ending with that API. 
We used FlowDroid~\cite{arzt2014flowdroid} to construct CGs and identify entry methods.
If an app contains calling contexts to an evolving dangerous API, we say the app uses the API.
This static analysis may produce \textbf{over-estimated} results with infeasible calling contexts.
In particular, the dangerous APIs may be called from third-party libraries (TPLs) but are not used by the host apps \cite{li2017libd,zhan2021atvhunter,zhan2020automated,ma2016libradar}.
To eliminate the influence of TPLs, we dropped those contexts whose entry methods lie in packages that are different from the app's package ID~\cite{wei2019pivot}.
This strict condition may filter out more contexts than necessary (\eg, when package names are obfuscated), hence producing \textbf{under-estimated} results.
To answer RQ2, we analyzed both the over-estimated and under-estimated results.


\subsubsection{Results}

\begin{table}[t]
\caption{Top-10 commonly used evolving dangerous APIs}
\label{tab:rq2}
\vspace{-1em}
\resizebox{\columnwidth}{!}{
\begin{tabular}{llrlr}

\toprule

Rank & Dangerous API &  Used apps & \begin{tabular}[c]{c}Dangerous API\\\footnotesize{(AppId-only)}\end{tabular} & \begin{tabular}[c]{c}Used apps\\\footnotesize{(AppId-only)}\end{tabular} \\

\midrule

1  &         {\mytt getDeviceId()} &       4,020 &         {\mytt getDeviceId()} &        592 \\
2  &           {\mytt getSerial} &       3,922 &      {\mytt getLine1Number} &        163 \\
3  &  {\mytt setRequireOriginal} &       2,305 &         {\mytt getAccounts} &        157 \\
4  &     {\mytt getCellLocation} &       1,980 &     {\mytt getSubscriberId} &        139 \\
5  &      {\mytt getLine1Number} &       1,792 &  {\mytt onCallStateChanged} &         91 \\
6  &     {\mytt getSubscriberId} &       1,714 &  {\mytt getSimSerialNumber} &         90 \\
7  &  {\mytt getSimSerialNumber} &       1,257 &           {\mytt getSerial} &         72 \\
8  &             {\mytt getImei} &       1,245 &             {\mytt getImei} &         59 \\
9  &         {\mytt getAccounts} &        968 &     {\mytt getCellLocation} &         54 \\
10 &         {\mytt getDeviceId(int)} &        913 &        {\mytt getGpsStatus} &         44 \\


\bottomrule

\end{tabular}

}
\end{table}

Table \ref{tab:rq2} lists the top-10 commonly used evolving dangerous APIs.
The left part presents the \textbf{over-estimated results}.
Under this setting, the API {\mytt getDeviceId()} is used by 4,020 apps, ranking at the first place.
This API requires a dangerous permission {\mytt READ\_PHONE\_STATE} in API level 23 to 28.
Since 29, it should be used with the signature permission {\mytt READ\_PRIVILEGED\_PHONE\_STATE}, and thus can no longer be used by general apps.
As described in the documentation \cite{WEB:android/getdeviceid}, for apps targeting 29 or above, calling this API will result in a {\mytt SecurityException}.
In our dataset, we found that each app has a median number of three calling contexts that can reach this API. If developers do not carefully deal with the permission changes in API level 29, the affected apps may encounter crashes at runtime.
In total, we observed that 69 out of the 188 evolving dangerous APIs are used in our dataset and 8,166 (61.2\%) of the 13,352 apps use at least one of these APIs.

The right part of Table \ref{tab:rq2} presents the \textbf{under-estimated results}.
Even with our strict filtering condition, we still found that 5,387 apps invoke dangerous APIs, while 1,051 (19.5\%) of them invoke at least one evolving dangerous API.
In addition, {\mytt getDeviceId()} is still the most commonly used evolving dangerous API.

In summary, we can see that {\bf a large number of apps may be affected by the evolution of permission specification}.

\subsection{RQ3: Permission Management Practices}
\label{ssec:rq3}

\subsubsection{Analysis Method}
\label{sssec:rq3-method}

Ideally, an app should always check the permission status before calling dangerous APIs, and request the permissions if they are not granted.
This can be accomplished by calling \textit{CHECK} and \textit{REQUEST} APIs synchronously before each dangerous API call, as shown in Figure \ref{fig:requestPermissions}.
The Android developer guide also suggests such a \textbf{\textit{synchronous permission management}}~\cite{WEB:android/request-permission}.
However, synchronous permission management can be impeded by many factors, \eg, developers' maintaining effort or user experience concern \cite{bonne2017exploring}.
In practice, developers may choose to implement permission management beyond the synchronous way.

\begin{figure}[t!]
    \includegraphics[width=\columnwidth]{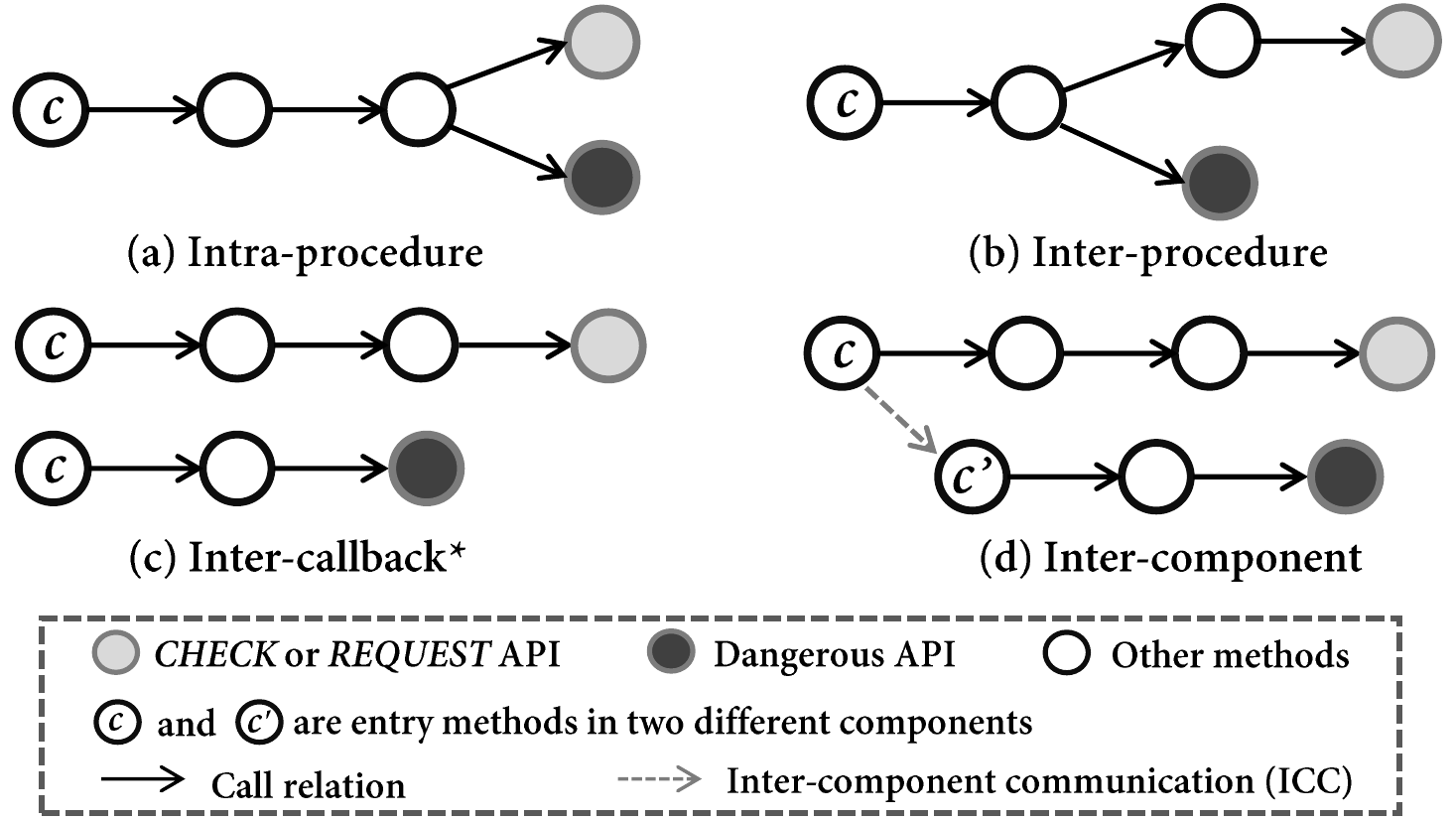}
    \vspace{-2em}
    \caption{Four types of permission managements ($^*$two entry methods are distinct callbacks in the same component)}
    \label{fig:practices}
    \vspace{-2.5ex}
\end{figure}

Figure \ref{fig:practices} illustrates four types of permission managements.
In Figure~\ref{fig:practices}(a)~and~\ref{fig:practices}(b), the dangerous API and the corresponding permission management API are invoked \textit{synchronously}.
The only difference between the two cases is that the \textit{CHECK}/\textit{REQUEST} is wrapped in another method call in Figure~\ref{fig:practices}(b).
It is noteworthy that we only consider \textit{CHECK} and \textit{REQUEST} APIs here because they are indispensable for runtime permission management (the other two types of APIs mentioned in \S~\ref{ssec:management} are optional).
Apart from the synchronous way, developers can call dangerous APIs and permission management APIs asynchronously in different callbacks \cite{yang2015static}, since Android programs are event-driven \cite{garcia2017automatic}.
For example, a dangerous API and the corresponding permission management API can be invoked from different callbacks in the same app component (Figure~\ref{fig:practices}(c)). It is also possible that dangerous API and the corresponding permission management API are invoked from the entry methods in two distinct components with inter-component communication (ICC) (Figure~~\ref{fig:practices}(d)).
We refer to the last two cases as \textit{\textbf{asynchronous permission management}}.

To study RQ3, we first extracted the calling contexts of dangerous APIs (\textit{dangerous contexts}) and permission management APIs (\textit{check/request contexts}) from each of the 13,352 apps.
Then we categorized these contexts into the above-mentioned four types of permission managements.
For example, a dangerous context and a check context will be considered as the \textit{intra-procedure} case, if both contexts have identical prefixes except their last nodes.
Due to page limit, we do not further elaborate on other cases.
It is worth explaining that a dangerous context may have multiple matching check contexts.
Imagining that there is another call to a \textit{CHECK} API inside {\mytt doLocationingActions()} in Figure \ref{fig:requestPermissions}, then the dangerous API called in {\mytt doLocationingActions()} would match both the intra-procedural and inter-procedural checks.
In this case, we would match the dangerous API with the closest one, according to the principle of locality \cite{xiong2017precise}.
Also note that the same dangerous API may have different permission mappings in different Android versions.
Here we only consider the mapping in the app's target SDK version, which is specified in its manifest file.

Our static analysis also needs to infer string values.
For example, when analyzing the code in~Figure \ref{fig:requestPermissions}, we should know the string arguments ({\mytt "ACCESS\_FINE\_LOCATION"}) of the \textit{CHECK} and \textit{REQUEST} API calls to match the permission management API calls with the dangerous API call in {\mytt doLocationingActions()}.
Such string analysis is difficult in general~\cite{christensen2003precise}.
However, we observed that developers often employ permission string literals (e.g., those defined in the class {\mytt android.Manifest.permission}) for permission managements, without performing complex string operations~\cite{li2015string}.
Hence, we modeled our string analysis task as a classic \textit{reaching-definition} dataflow analysis problem \cite{aho2007compilers} ($s$ is a statement; $\mathit{Pred}(s)$ returns the predecessor statements of $s$; $\mathit{gen}(s)$ and $\mathit{kill}(s)$ are the dataflow facts generated or killed by $s$, respectively):
\begin{align}
\mathit{IN}(s) &= \bigcup_{p\in \mathit{Pred}(s)}\mathit{OUT}(p) \label{eq:redef-in} \\
\mathit{OUT}(s) &= \mathit{gen}(s)\cup(\mathit{IN}(s)-\mathit{kill}(s)) \label{eq:redef-out}
\end{align}
When the iterative equation solving converges, the possible string values on a call site $c$ can be retrieved from the set $\mathit{IN}(c)$.
To avoid unnecessary computation, we only propagate dataflow facts related to string and string array variables.

Most Android-specific analyses, such as identifying app components and building intra-component CGs, rely on FlowDroid.
To analyze ICC, we used the ICC link extraction rules introduced in IccTA \cite{li2015iccta}.
We did not directly apply the IccTA tool as it relies on IC3, which is outdated and failed to analyze most of our apps.
To realize inter-procedural reaching definition analysis, we employed the IFDS solver Heros~\cite{bodden2012inter}.
To ensure that the permission usages are implemented in the apps rather than TPLs, we only considered the calls to dangerous and permission management APIs from the apps' packages. 2,402 apps in our dataset satisfy this condition.

\subsubsection{Results}

\begin{figure}[t!]
    \includegraphics[width=\columnwidth]{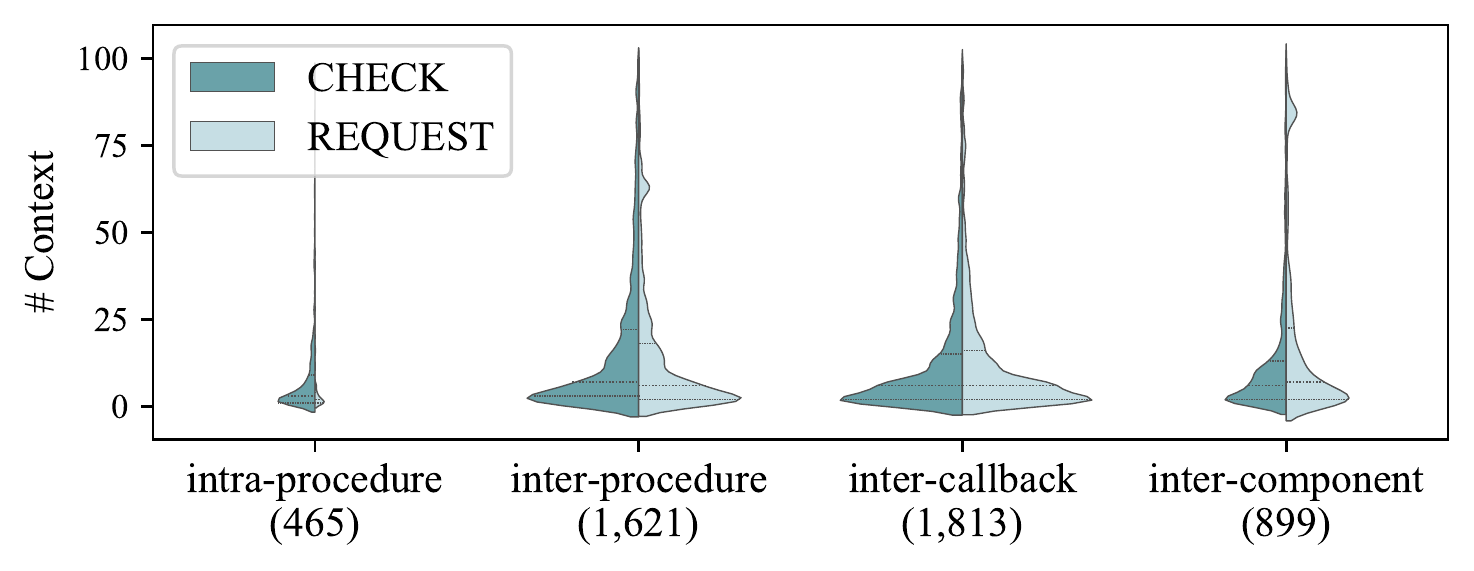}
    \vspace{-2em}
    \caption{Distribution of dangerous API calling contexts with respect to the four types of permission managements}
    \label{fig:rq3}
    \vspace{-2.5ex}
\end{figure}

Figure~\ref{fig:rq3} presents the dangerous API calling contexts with respect to the four types of permission managements using violin plots.
For better visualization, we excluded 34 apps with an extremely large number (over 100) of dangerous API calling contexts.
For ease of understanding, we explain the first violin plot. The left part of the violin has a median of three, the right part has a median of two, and the violin is labeled with 465.
This means that there are 465 apps that implement intra-procedural permission checks or requests.
Among these apps, a median number of three dangerous API calling contexts are protected by intra-procedural permission checks, while a median of two dangerous API calling contexts have intra-procedural permission requests.

Intuitively, the area of each violin reflects the adoption of the corresponding permission management practice.
For instance, the right part of the first violin is the smallest, meaning that only a few apps request permissions inside the same method where the dangerous APIs are called.
In fact, only 114 apps contribute to this part.
The intra-procedural permission checks are also the least common as shown in the left part.
In contrast, 1,621 apps adopt inter-procedural permission managements, which usually happen when developers wrap the \textit{CHECK}/\textit{REQUEST} API calls in self-defined methods for code reuse (e.g., in \textit{Syncthing} issue 1575~\cite{WEB:github/syncthing/1575}).

\textbf{Asynchronous permission managements are more common than synchronous permission managements}: According to the results, 1,813 of the 2,402 apps perform inter-callback permission management and 899 apps perform inter-component permission management.
In total, 2,065 out of the 2,402 (86.0\%) apps adopt asynchronous permission management.
We further investigated the calling contexts of asynchronous \textit{CHECK}s and \textit{REQUEST}s, and found that most of their entry methods are {\mytt onClick()} and {\mytt onCreate()}.
There is a large number of \textit{REQUEST} calls starting from {\mytt onStart()}, in which UI elements are suggested to be drawn~\cite{WEB:android/activity-onstart}.
In these scenarios, users usually receive permission request dialogs after they click a button that performs sensitive operations, or switch to a new activity that requires dangerous permissions.
Note that these results were obtained following the under-estimation setting discussed in \S~\ref{sssec:method}. We also did the same analysis following the over-estimation setting and observed similar context distributions. Due to page limit, we omit the details.

\textbf{Discussions.}
Asynchronous permission management can avoid redundant \textit{CHECK}s and \textit{REQUEST}s.
However, it brings challenges to developers as they should examine both the dangerous API call sites and all related app components to see whether the required permissions are already requested and granted.
Moreover, they should carefully handle the case when the users revoke a particular permission required by the dangerous API, since asynchronous permission requests may be unavailable in this scenario.

Asynchronous permission management practices also complicate ARP bug detection.
A conservative detector that considers an app buggy whenever a synchronous \textit{CHECK} or \textit{REQUEST} is missing at a dangerous API call site (\eg, \textsc{RevDroid}~\cite{fang2016revdroid}) may produce many false alarms, if asynchronous permission management can already protect the dangerous API call.
Furthermore, as users may revoke permissions between two asynchronous events, how to determine whether a \textit{CHECK} can protect a subsequent dangerous API call is also a technical challenge for static bug detectors.
As we will see in \S~\ref{ssec:case}, asynchronous permission managements can lead to both safe permission usages and app crashes, which cannot be distinguished by existing tools.

\section{\tooln}
\label{sec:aper}

\subsection{Overview}

Our empirical study reveals that a large number of real-world apps perform asynchronous permission managements, which complicates permission checking and may lead to subtle ARP bugs.
Moreover, compatibility issues may arise due to the active changes of the mappings between dangerous APIs and their required dangerous permissions (\textit{API-DP mappings}).
To ease subsequent discussion, we first define two common types of \textit{ARP bugs}:
\begin{itemize}
\item \textbf{Type-1 (Missing Permission Check)}: A dangerous API is called without a permission check on the target Android version.
\item \textbf{Type-2 (Incompatible Permission Usage)}: A dangerous API can be called on incompatible platforms, or the evolution of permission specification is not fully handled.
\end{itemize}
According to \cite{wang2021runtime}, these two types of bugs correspond to the non-library-interfered ARP issues except those caused by device manufacturers' customization, which account for 60.8\% of their studied real ARP issues.
Existing tools cannot effectively detect these two types of common ARP bugs in Android apps.
This motivates us to design a new approach to detect these bugs. 

\begin{figure}[t!]
    \centering
    \includegraphics[width=0.95\columnwidth]{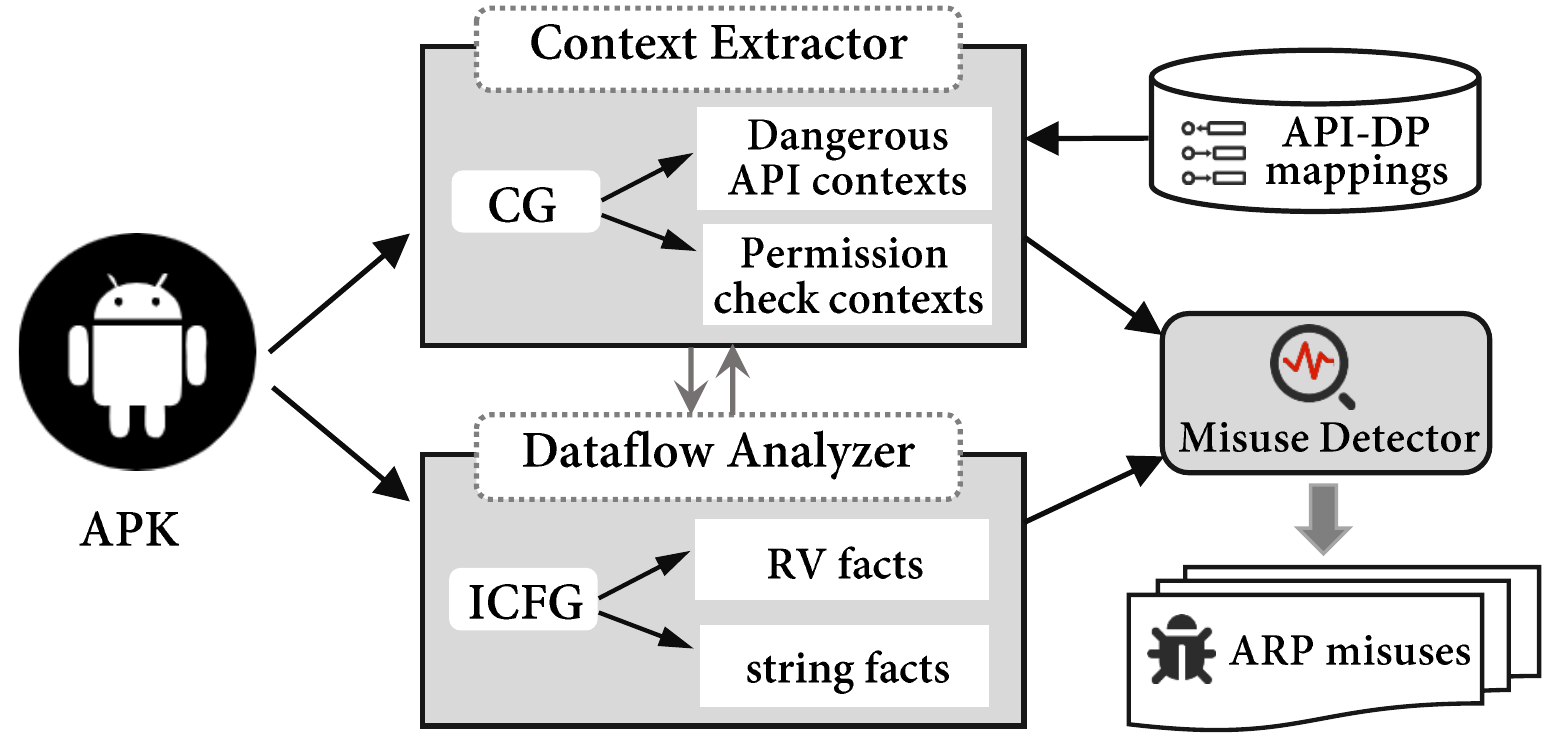}
    \caption{Overview of \tooln}
    \label{fig:aper}
\end{figure}

Figure~\ref{fig:aper} shows the overview of our tool, \tooln.
It takes an Android application package (APK) file and the API-DP mappings as input, and outputs the calling contexts of dangerous APIs that have potential Type-1 or Type-2 bugs.
Its analysis procedure consists of three components:

\begin{enumerate}

\item The \textbf{Context Extractor} traverses the CG of an app and performs backward analysis from the call sites of dangerous APIs and \textit{CHECK} APIs to extract all possible calling contexts of these APIs.
Similar to our approach in the empirical study, a dangerous context will be matched with \textit{CHECK} contexts (Figure~\ref{fig:practices}) according to the API-DP mappings.
The only difference here is that the mappings for all Android versions will be used, in order to detect Type-2 bugs.
The processed calling contexts will be analyzed by the Misuse Detector to locate ARP bugs.

\item In order to know which permissions are checked at each call site of the \textit{CHECK} API, the \textbf{Dataflow Analyzer}
conducts a reaching definition analysis on the inter-procedural control-flow graph (ICFG) of the app to infer the possible string values passed to the \textit{CHECK} API (\S~\ref{sssec:rq3-method}).
This helps Context Extractor to match a dangerous context with the \textit{CHECK} contexts that possibly protect the dangerous API calls.
To detect Type-2 bugs, the Dataflow Analyzer also analyzes the runtime version (RV) checks that guard dangerous API calls to infer reachable RVs on which the dangerous APIs can be invoked.

\item The \textbf{Misuse Detector} leverages the API calling contexts and dataflow facts generated by the other components to locate bugs.
The idea is: A dangerous API context without proper permission checks may contain Type-1 bugs; An evolving dangerous API context without proper RV checks may contain Type-2 bugs.

\end{enumerate}

It is worth noting that \tool does not analyze the usages of the \textit{REQUEST} APIs, as the request process is asynchronous and cannot guard the calls to dangerous APIs. Nonetheless, the Misuse Detector will consider the dangerous API calls within the \textit{HANDLE} callback (\S~\ref{ssec:misuse}), which handles the request results.
Most of the analyses performed by \tool have been detailed in our empirical study (\S~\ref{sssec:rq3-method}). In the following, we explain how \tool analyzes RV checks and detects Type-1/Type-2 bugs.

\subsection{Analyzing Reachable Runtime Versions}
\label{ssec:rvs}

\begin{figure}[t]
    \resizebox{\columnwidth}{!}{ 
\begin{lstlisting}[language=java]
if(Build.VERSION.SDK_INT >= 29 &&
        checkSelfPermission(ACCESS_FINE_LOCATION) != GRANTED){
    requestPermissions(new String[]{ACCESS_FINE_LOCATION});
    return;
}
p2pManager.createGroup(...); // evolving dangerous API
\end{lstlisting} }
    \vspace{-1em}
    \caption{An Example of Handling API Evolution}
    \label{fig:RepeaterManager}
    \vspace{-2ex}
\end{figure}

Figure~\ref{fig:RepeaterManager} shows how the developers of a popular app, \textit{VPNHotSpot}~\cite{WEB:github/VPNHotspot}, handled the evolution of a dangerous API.
Starting from API level 29, the API {\mytt WifiP2pManager.createGroup()} requires the dangerous permission {\mytt ACCESS\_FINE\_LOCATION}, while it does not require permissions in the previous API levels.
As the example shows, to invoke the dangerous API on compatible Android versions, the app first checks the value of the field {\mytt Build.VERSION.SDK\_INT}, which is widely used to handle incompatible API usages~\cite{li2018cid}.

To detect incompatible permission usages, \tool needs to infer the possible RVs on which a dangerous API can be invoked.
For this purpose, \tool performs \textit{dominator analysis}~\cite{song2019servdroid}.
In a flow graph, a node $d$ \textit{dominates} another node $n$ if $d$ exists on every path starting from the entry node to $n$ (\ie, to reach $n$, its dominator $d$ must be gone through).
In an Android app, if a dangerous API call site is dominated by some RV checks, these checks then constrain the RVs on which the API can be invoked. We call such RVs \textit{reachable RVs}.

Algorithm~\ref{algo:rv-analysis} explains how \tool analyzes reachable RVs.
Given a dangerous context, \tool analyzes all methods invoked prior to the dangerous API (lines 2-9) to look for dominating RV checks (line 6).
A statement is an RV check if it is a conditional statement and the condition refers to the constant field {\mytt Build.VERSION.SDK\_INT}\footnote{This approach may miss some runtime version checks. However, such cases are rare in practice \cite{li2018cid}.}.
When an RV check is found, \tool will solve the constraint involved in the condition to find all reachable RVs (line 7).
If there exist multiple dominating RV checks, \tool will compute the conjunction of the reachable RVs constrained by each check (line 8).

\begin{algorithm}[t!]
\caption{Analyzing reachable RVs}
\label{algo:rv-analysis}
\DontPrintSemicolon
\SetInd{0.4em}{0.9em}
\SetKwProg{Fn}{Function}{:}{}

\KwIn{Dangerous API calling context $p_1\shortrightarrow p_2\shortrightarrow\cdots\shortrightarrow p_\textrm{n}$\\
\hspace*{3em}($p_i$ are call sites, and $p_\textrm{n}$ calls a dangerous API)}
\KwOut{Set of reachable RVs $\subseteq$ $\{23,...,\textrm{LAV}^\dagger\}$}
\BlankLine
{\it reachableRV} $\shortleftarrow$ $\{23,...,\textrm{LAV}\}$\;
\For{$~i \shortleftarrow 1\dots${\rm n-1}$~~$}{
    {\it m} $\shortleftarrow$ {\sc getBodyOfCalledMethod}($p_\textrm{i}$)\;
    {\it dominators} $\shortleftarrow$ {\it m}.{\sc findDominatorsOf}($p_\textrm{i+1}$)\;
    \For{$~${\it stmt} $\in$ {\it dominators}$~$}{
        \If{{\sc isRvCheck}({\it stmt})}{                                \label{ln:isRvCondition}
            {\it rvs} $\shortleftarrow$ {\sc solveAllSatisfiable}({\it stmt})\;
            {\it reachableRV} $\shortleftarrow$ {\it reachableRV}$\ \cap\ ${\it rvs}\;
        }
    }
}
\Return{reachableRV}

\algorithmfootnote{$^\dagger$: ``LAV'' stands for the Latest Android Version.}
\end{algorithm}

It should be pointed out that, RV checks may not necessarily dominate a dangerous API call site in order to protect it.
Consider two separate RV checks in two parallel program branches and both branches can lead to the same dangerous API call.
In such a case, neither check dominates the dangerous API call, although it is guaranteed to be called safely.
To reduce algorithmic complexity, \tool does not take such unusual cases into account.

\subsection{Detecting ARP Bugs}
\label{ssec:misuse}

The Misuse Detector detects ARP bugs by analyzing each dangerous context and two other pieces of information:
1) a set of matching \textit{CHECK} contexts provided by the Context Extractor, and 2) all reachable RVs provided by the Dataflow Analyzer.
In the following, we describe how \tool detects Type-1 and Type-2 bugs in detail.

\subsubsection{Detecting Type-1 Bugs}

A Type-1 bug occurs when a dangerous API is called without permission checks on the target Android version.
In the most trivial case, a Type-1 bug can be reported when the set of matching \textit{CHECK} contexts is empty.
If the set is not empty, we need to analyze whether the \textit{CHECK} contexts could safely protect the dangerous context.
There are three cases to consider:

\begin{enumerate}
\item If the \textit{CHECK} API and the dangerous API are synchronously called, \tool will examine whether the former's call site dominates the latter's on the ICFG and whether the dangerous API is reachable from the \textit{CHECK} API's positive branch.
\item If the \textit{CHECK} and dangerous APIs are called in different callbacks of an app component, \tool will examine whether the \textit{CHECK} context's entry method precedes the dangerous context's (\eg, {\mytt onCreate()} must precede {\mytt onStart()}, {\mytt onResume()} must precede {\mytt onClick()}).
If a dangerous API is called in callback A while a \textit{CHECK} is called in callback B that must precede A, such a permission check is safe.
To analyze the execution order of callbacks, we used the temporal constraints defined by Liu \etal~\cite{liu2016understanding} to distinguish erroneous asynchronous permission checks from safe cases.
\item If the \textit{CHECK} API is called in a component $c$, while the dangerous API is called in another component $c'$, \tool will examine whether the \textit{CHECK} API's call site (more specifically, its positive branch) in $c$ dominates the call site of the ICC method that launches $c'$.
\end{enumerate}

In each case, if a dangerous context has no dominating \textit{CHECK}s, \tool will report a Type-1 bug.
To avoid false alarms, \tool will not report bugs in two cases:
1) The dangerous API call is wrapped in a {\mytt try-catch} block that handles the {\mytt SecurityException}.
Some developers may use such a workaround to avoid app crashes, instead of explicitly performing permission checks before calling dangerous APIs.
2) Inside the \textit{HANDLE} callback, the dangerous API is called after checking the permission request results.
Since the permission request results are passed as a parameter to this callback, developers can check the parameter value to learn the permission status and invoke dangerous APIs when the permission is granted.

\subsubsection{Detecting Type-2 Bugs}

A Type-2 bug can occur when a dangerous API is called on incompatible Android versions (calling new APIs on old platforms, or removed APIs on new platforms), or the permissions required on different versions are not fully handled.
\tool detects Type-2 bugs by examining whether a dangerous API can be safely called on all reachable RVs, except the app's target SDK version, which is already analyzed when detecting Type-1 bugs.
Given a dangerous context, with a corresponding reachable RV $v$, if the dangerous API does not exist in the API-DP mappings of the Android version $v$, then a Type-2 bug can be reported. This can happen if the API has been deleted or not yet introduced.
If the API exists in the mappings, then \tool will analyze whether the API is invoked with a dominating check of the corresponding permission required on the Android version $v$, which is essentially the same as detecting Type-1 bugs. \tool will report a Type-2 bug when there are no dominating permission checks.

\section{Evaluation}
\label{sec:evalution}

Our evaluation aims to answer two research questions:
\begin{itemize}
\item \textbf{RQ4 (Effectiveness of \tooln)}: How effective is \tool in detecting ARP bugs, compared with the existing tools?
\item \textbf{RQ5 (Usefulness of \tooln)}: Can \tool detect unknown ARP bugs in real-world apps and help developers diagnose them?
\end{itemize}
In the following, we present our experiments and analyze the results in detail. We also discuss some real ARP bugs detected by \tooln.

\subsection{RQ4: Effectiveness of \tool}
\label{ssec:rq4}

\subsubsection{Constructing Benchmark}

To study RQ4, our evaluation subjects should contain both Type-1 and Type-2 bugs.
More importantly, to understand how the tools report false alarms, we need subjects that use dangerous APIs correctly.
For this purpose, we constructed a benchmark, \benchn, leveraging real ARP bugs in open-source apps and their patches. 
Specifically, we collected projects from GitHub \cite{WEB:github} that have both
1) issues related to \textit{CHECK} APIs and dangerous APIs (to identify ARP bugs), and
2) commits or pull requests that fix the issues (to locate the bug fixes).
We found 61 such projects with a total of 71 ARP issues, which correspond to 71 ARP bugs.
For each bug, we then applied the following procedure on both the buggy version and the patched version:
1) Locating the related dangerous API;
2) Finding the API's corresponding permission management code via manual inspection;
3) Removing irrelevant app classes, methods, attributes, and TPLs, until a minimal compilable APK remains. 
For the ease of experiments, each APK has only one dangerous API, and thus a buggy APK contains only one ARP bug. 
If we could not build an APK, we simply discarded that bug.
Finally, 60 buggy APKs were successfully built, among which 35 contain Type-1 bugs and 25 contain Type-2 bugs. Correspondingly, each buggy APK has its patched version. 

\subsubsection{Baselines and Metrics}

To the best of our knowledge, there are three available static analysis tools that can find ARP bugs:
\begin{itemize}
\item \textsc{Lint}~\cite{WEB:android/lint} is a built-in checker in Android Studio, which is the official IDE for developing Android apps.
It can report missing permission checks and RV checks, and we treat these two types of warnings as Type-1 and Type-2 bugs, respectively. 
\item \textsc{ARPDroid}~\cite{dilhara2018automated} is an academic tool for automatic runtime permission management. If it inserts permission management statements on any calling context of a dangerous API, we consider that it detects a Type-1 bug.
\item \textsc{RevDroid}~\cite{fang2016revdroid} is also an academic tool, which can detect unhandled permission revocation on a dangerous API call, and we treat its warnings as Type-1 bugs.
\end{itemize}
For fair comparisons, we did not consider testing-based tools (\eg, \textsc{PATDroid} \cite{sadeghi2017patdroid} or \textsc{SetDroid} \cite{sun2021understanding}) because their performances heavily rely on the underlying tests' coverages.

To answer RQ4, we applied \tool and the three tools on \bench and compared their performance using the following metrics:
\begin{itemize}
\item True positives (\textit{TP}): \# buggy versions that have warnings
\item True negatives (\textit{TN}): \# patched versions that have no warnings
\item False positives (\textit{FP}): \# patched versions that have warnings
\item False negatives (\textit{FN}): \# buggy versions that have no warnings
\end{itemize}
Based on them, we can further calculate \underline{P}recision ($\frac{\mathit{TP}}{\mathit{TP}+\mathit{FP}}$), \underline{R}ecall ($\mathit{\frac{TP}{TP + FN}}$), and $F_1$-score ($\mathit{\frac{2\cdot P\cdot R}{P+R}}$) to measure their effectiveness.

\subsubsection{Results and Analyses}

\begin{table}[t]
\caption{Comparison results with the existing tools}
\label{tab:rq4}
\vspace{-1em}
\resizebox{\columnwidth}{!}{
\begin{threeparttable}

\begin{tabular}{@{}ccccccc@{}}
\toprule
\multirow{2}{*}{} & \multicolumn{4}{c}{\textbf{Type-1}}          & \multicolumn{2}{c}{\textbf{Type-2}} \\ 
\cmidrule(lr){2-5}
\cmidrule(l){6-7}
                  & \textsc{Lint}  & \textsc{ARPDroid} & \textsc{RevDroid} & \textsc{Aper}  & \textsc{Lint}         & \textsc{Aper}        \\ 
\midrule
$\mathit{TP}$                & 16    & 13       & 15       & 26       & 14           & 23          \\
$\mathit{TN}$                & 22    & 22       & 26       & 32       & 15           & 19          \\
$\mathit{FP}$                & 13    & 13       & 6        & 3        & 10           & 6           \\
$\mathit{FN}$                & 19    & 22       & 17       & 9        & 11           & 2           \\
Failed            & -     & -        & 6$^*$        & -        & -           & -           \\
Precision (\%)    & 55.17 & 50.00    & 71.43    & \textbf{89.66}  & 58.33    & \textbf{79.31}       \\
Recall (\%)       & 45.71 & 37.14    & 46.88   & \textbf{74.29} & 56.00        & \textbf{92.00}       \\
$\mathit{F_1}$-score (\%)    & 50.00  &  42.62  & 56.61   &   \textbf{81.25} & 57.14     &  \textbf{85.19}  \\
\bottomrule
\end{tabular}

\begin{tablenotes}
    \item $^*$\textsc{RevDroid} fails on both versions of the three subjects.
\end{tablenotes}

\end{threeparttable}
}
\end{table}

Table~\ref{tab:rq4} shows the comparison results.
We can see that \tool outperforms the baselines on all metrics.
For Type-1 bugs, \tool achieves an $F_1$-score of 81.25\%, with an improvement of 43.5\% ($=$$(81.25$-$56.61)/56.61$) over \textsc{RevDroid}, which performed the best among the baselines.
As for Type-2 bugs, the improvement of \tool over \textsc{Lint}, which is the only baseline tool that supports detecting Type-2 bugs, on $F_1$-score is 49.1\%.
On average, \tool outperforms existing tools by 46.3\% ($=$$(43.5$+$49.1)/2$) on $F_1$-score, indicating its effectiveness on ARP bug detection.

For Type-1 bugs, the baseline tools suffer from low precisions.
We manually investigated these tools and figured out the main reasons. First, \textsc{Lint} performs flow-insensitive analysis and only detects the presence of \textit{CHECK} without considering the API-DP mappings or the return value of \textit{CHECK}.
For \textsc{ARPDroid}, we found that it only inserts \textit{CHECK}s and \textit{REQUEST}s without properly dealing with the existing calls to these APIs.
The best baseline, \textsc{RevDroid}, leverages FlowDroid to generate analysis entry points and cannot handle ICC well~\cite{qiu2018analyzing}.
As we will see in \S~\ref{ssec:case}, only performing intra-component analysis may produce many false alarms.
In comparison, \tool explicitly models both synchronous and asynchronous permission checks, which greatly improves the precision.

The baseline tools have low recalls on both types of bugs.
The primary reason is the incompleteness of the API-DP mappings, which is also a major reason for \tooln's low recall on Type-1 bugs.
For example, the API {\mytt Camera.open()} does not specify any permission in the documentation, and thus is not covered in our mappings.
However, in practice, apps would request the user to grant {\mytt CAMERA} permission before invoking it~\cite{WEB:stackoverflow/permission/camera}.
Another reason is that some issues in our benchmark are related to content providers. Currently, neither the baselines nor \tool can analyze the behavior of content providers. We will address the limitations of \tool in the future.

\subsection{RQ5: Usefulness of \tool}
\label{ssec:rq5}


\subsubsection{Collecting Real-World Apps}

To answer RQ5, we collected 214 apps from F-Droid \cite{WEB:f-droid}, a widely used open-source app catalogue, to see whether \tool can detect ARP bugs from real-world apps.
These apps were selected following three criteria:
1) invoking at least one dangerous API,
2) containing identifiable \textit{CHECK}s or \textit{REQUEST}s,
and
3) having commit records in the recent three years in their code repository.
We only performed experiments on open-source apps because we need the source code to verify the detected bugs, manually reproduce app crashes, and communicate with the app developers to investigate whether \tool is useful.
It should be noted that \tool can also help detect bugs in closed-source apps.

\subsubsection{Study Method}

We ran \tool on the 214 apps, and manually verified its reported ARP bugs.
We first investigated the reported contexts to see if Aper was functionally correct.
Then, we tried to reproduce abnormal behaviors based on the given context.
For example, if \tool reports a Type-1 bug on {\mytt BackupActivity}, we would launch the app, go to its backup page, then revoke the concerned permission(s) in the system settings, and go back to the app.
Typically the app would crash (or behave abnormally) after we re-entering it.
However, if we could not find the desired page (\eg, the {\mytt BackupActivity}) after thoroughly exploring the app, we treated this case as unverified.
As for Type-2 bugs, we would verify them under a similar procedure with Type-1 bugs, but on multiple Android devices.
Our main focus was whether their runtime behaviors were consistent across different Android versions.
We used 13 Android devices, most of which run the stock Android systems with API levels ranging from 23 to 30.

Once we verified an ARP bug, we reported it to the developers via the issue tracking system links provided on F-Droid.
To help developers diagnose the bugs, we recorded videos and appended them to the bug reports, if we could reproduce them.
For some bugs, we also sent our suggested patches to the developers to fix them.

\emph{\textbf{Ethical Considerations}}. 
To avoid spamming the open-source community and the developers, we reported an issue only when it could be reproduced on at least three different Android devices, and submitted pull requests only after we had thoroughly tested the patched code.
All issue reports and pull requests were submitted in compliance with the projects' contributing guidelines and licenses. 

\subsubsection{Results}

\begin{table}[t]
\caption{Summary of detected ARP bugs}
\label{tab:rq5}
\vspace{-1em}
\resizebox{0.98\columnwidth}{!}{
\begin{tabular}{@{}ccccccc@{}}
\toprule
       & Detected & Verified & Reported & Confirmed & Fixed & \# Videos \\ \midrule
Type-1 & 66       & 23       & 20       & 12        & 3     & 21        \\
Type-2 & 18       & 11       & 10       & 5         & 4     & 3         \\ \bottomrule
\end{tabular}

}
\end{table}

Table \ref{tab:rq5} summarizes the ARP bugs detected by \tooln.
Among the 84 detected ARP bugs, we identified 23 true Type-1 bugs and 11 true Type-2 bugs.
We reported 30 of the 34 true bugs.
We did not report the other four bugs because the issue tracking systems of the corresponding apps were closed at the time when we conducted the experiments.
Nonetheless, we still put the bug reproduction videos on our project site.
At the time of our paper acceptance, 17 of our reported bugs have been confirmed and seven bugs have been fixed.
Specifically, three bugs were fixed by our suggested patches.
Interestingly, two of the fixed bugs are in TPLs: \textit{ACRA} \cite{WEB:github/acra} and \textit{KAHelpers} \cite{WEB:github/kahelpers}.
Both of them are popular on GitHub (with 5.7k and 520 stars, respectively).
During our manual bug verification, we found that \tool reported ARP bugs in them when analyzing the host apps.
After analyzing the root causes of the bugs, we decided to report them to the library developers, rather than the app developers.
Such a case will be discussed in \S~\ref{sssec:case-3}.
In total, 24 videos were successfully recorded. 
Among them, 20 videos recorded test cases that trigger app crashes, two were about program stuck and the remaining two were inconsistent behaviors on different Android devices.
Each test case was constructed by us in typically less than ten minutes, after examining \tooln's output and the app's source code.
These results show that \tool can detect real ARP bugs and produce useful debugging information.

21 detected bugs are false alarms.
A majority of them are caused by specific API usages.
For example, 13 bugs are related to the dangerous API {\mytt getExternalStorageDirectory()}, which returns a directory path in the external storage.
A recommended way to examine this path's readability is to check the {\mytt READ\_EXTERNAL\_STORAGE} permission \cite{WEB:stackoverflow/externalstorage}.
However, developers may also use other APIs for the same purpose, such as {\mytt file.canRead()} in JDK.
In this situation, \tool will report a Type-1 bug for lack of \textit{CHECK} APIs, although the file access is safe.
We failed to verify the remaining bugs because triggering them requires complicated setups or interactions (\eg, triggering a bug in Kore~\cite{WEB:github/kore} requires a connected Kodi player). 
It is worth noting that these unverified bugs are not necessarily false alarms.

\subsection{Case Studies}
\label{ssec:case}

In this subsection, we discuss several typical cases observed during our study to facilitate the design of future tools.

\subsubsection{A Type-1 ARP Bug}

\begin{figure}[t!]
    \centering
    \includegraphics[width=0.95\columnwidth]{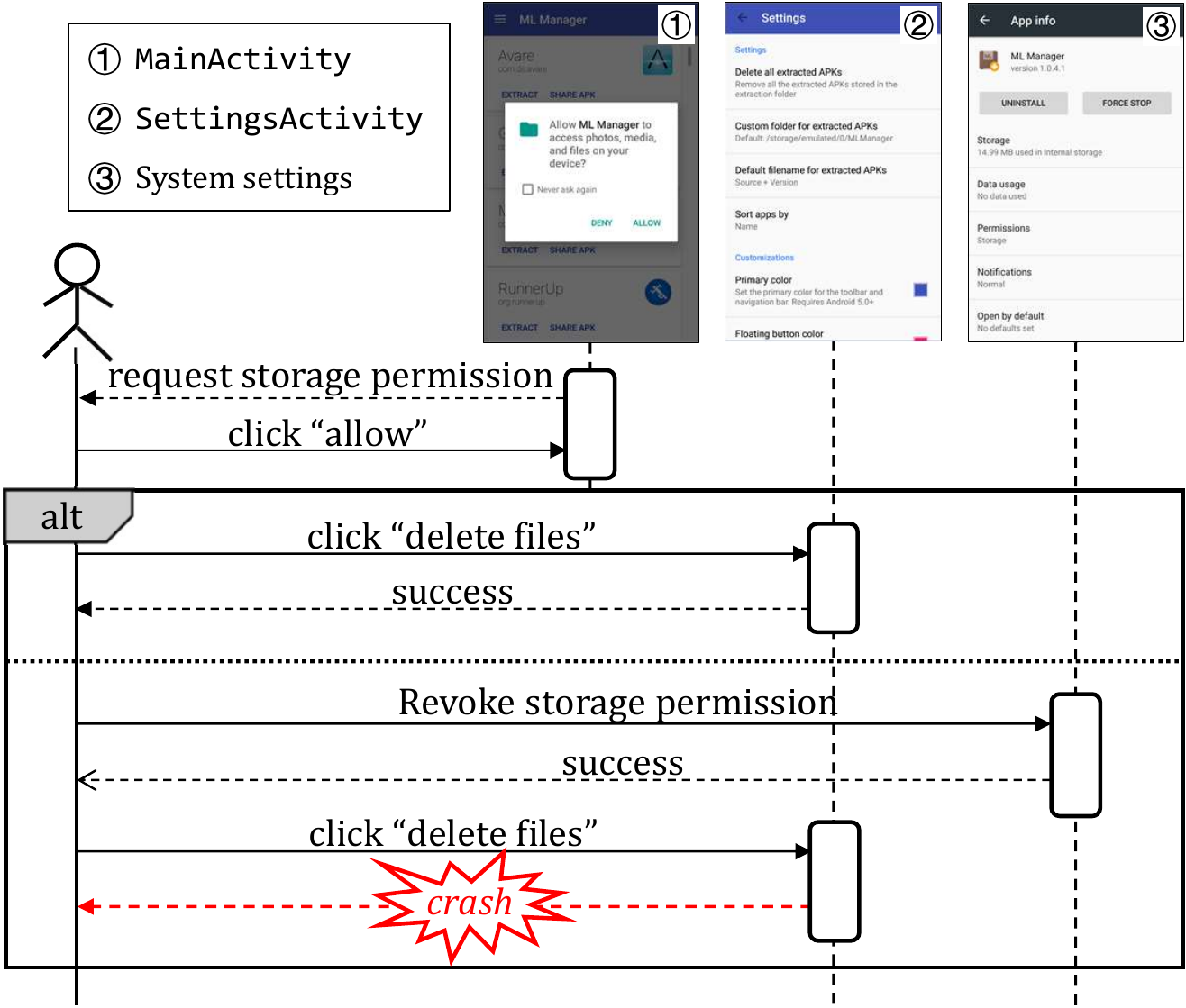}
    \vspace{-1em}
    \caption{An inter-component Type-1 bug in ML Manager}
    \label{fig:case1}
    \vspace{-2ex}
\end{figure}

Figure \ref{fig:case1} shows a Type-1 bug caused by inter-component permission usage.
It was found in \textit{ML Manager} \cite{WEB:github/MLManager}, an open-source app manager with 100K+ installs on Google Play.
Normally, the user grants the storage permission in the app's  {\mytt MainActivity}, and then the app can delete cached files when the user operates in the {\mytt SettingsActivity}.
However, because the latter component does not involve \textit{CHECK}s or \textit{REQUEST}s, once the user revokes the storage permission after entering it, the app can no longer access the device's external storage.
In such a scenario, deleting files will result in a crash.
This bug was fixed in the app's 2.5.2 version.
Developers added \textit{CHECK} in the button click listener to avoid unprotected storage access.
We found that three out of the four tools could detect this bug, except \textsc{Lint}, since the app has \textit{CHECK} statements in other unrelated components.

\subsubsection{A Safe Inter-Component ARP Usage}

\begin{figure}[t]
\begin{lstlisting}[language=java]
void onCreate(Bundle bundle){
  // ...
  if(isSdCardPermissionNotGranted())
    requestSdCardPermissions();
  else {
    intent = new Intent(DirectoryPickerActivity);
    startActivity(intent);
  }
}
\end{lstlisting}
    \vspace{-1em}
    \caption{A safe inter-component ARP usage in Tricky Tripper (in class {\mytt SaveToSdCardActivity}, simplified)}
    \label{fig:case2}
    \vspace{-2ex}
\end{figure}

Static analyzers may report false alarms on safe inter-component ARP usages.
Figure~\ref{fig:case2} shows such an example from the app \textit{Tricky Tripper}~\cite{WEB:github/trickytripper}.
The {\mytt DirectoryPickerActivity} displays the file directory and thus needs storage permission.
This activity can only be launched when the permission is granted, so the usage is safe.
However, both \textsc{ARPDroid} and \textsc{RevDroid} would report the lack of permission management in {\mytt DirectoryPickerActivity}.
In contrast, \tool can avoid this $\mathit{FP}$ by resolving the ICC between the two components.

\subsubsection{A Type-2 Bug in TPLs}
\label{sssec:case-3}

\begin{figure}[t]
    \resizebox{\columnwidth}{!}{ 
\begin{lstlisting}[language=diff]
  override fun collect(...) {
    // ...
-   target.put(ReportField.DEVICE_ID, telephonyManager.deviceId)
+   val deviceId = if(SDK_INT<=28) telephonyManager.deviceId else null
+   target.put(ReportField.DEVICE_ID, deviceId)
  }
\end{lstlisting} }
    \vspace{-1em}
    \caption{Pull request \#890 in ACRA library (simplified)}
    \label{fig:case3}
    \vspace{-2ex}
\end{figure}

As we discussed, some bugs can occur in TPLs.
Instead of blaming the host apps, we reported these issues to the library developers.
Figure \ref{fig:case3} shows our pull request in the library \textit{ACRA}, written in Kotlin language, which fixes the issue caused by the evolution of the dangerous API {\mytt getDeviceId()}.
After compilation, the expression {\mytt telephonyManager.deviceId} is transformed into this dangerous API.
The API {\mytt getDeviceId()} is commonly used in real-world apps, as shown in Table~\ref{tab:rq2}.
Since API level 29, many apps may throw a {\mytt SecurityException} when invoking this API because it starts to require a signature permission, which is not available to general apps.
To avoid app crashes, we enforce the API to be called under API level 29, and the API will return a {\mytt null} value on newer Android versions.
This patch has been approved by the project maintainer, and will be merged to the next release of the library.

\subsubsection{An Intended Permission Misuse}

In fact, not all detected bugs are considered harmful from the developers' perspectives.
For example, the API {\mytt getConfiguredNetworks()} starts to require fine-location permission since API level 29.
In \textit{NetGuard}~\cite{WEB:github/netguard}, an Internet firewall app with over five million installs, \tool reported that this API call lacks permission checks on API level 29 and above.
The app developer confirmed this is true~\cite{WEB:netguard/issue}.
However, for privacy concerns, they do not want their app to request the location permission.
They would rather disable the related features.

\section{Threats to Validity}
\label{sec:threats}

The validity of our study results is subject to the following threats:

\noindent$\bullet$ \textbf{Incomplete API-permission mappings.} We built the mappings between APIs and permissions via analyzing the annotations and Javadocs in the source code of the Android framework.
We did not directly use mappings from previous work due to the following reasons:
1) As the Android platform constantly evolves, existing mappings can quickly become outdated. For example, Arcade \cite{aafer2018precise} only released the mappings up to API level 25, while the most recent work, Dynamo \cite{dawoud2021bringing}, only released mappings for two API levels, 23 and 29.
2) The mapping extraction tools are unavailable or difficult to use.
3) The existing mappings are either imprecise or incomplete, as pointed out by \cite{dawoud2021bringing}. An example is PScout \cite{au2012pscout}, which was claimed to contain at most 7\% incorrect mappings.
In contrast, although our approach may extract incomplete mappings, they are precise and, more importantly, up-to-date with the latest Android version.

\noindent$\bullet$ \textbf{False positives in the extracted contexts.} Extracting all dangerous API calling contexts to understand the ARP practices may lead to false positives. To mitigate the threat, we proposed the under-estimation setting which only considers entry methods within the app's package. We also performed manual validation by sampling apps from F-Droid and inspecting their extracted contexts to ensure the precision of our analysis.

\noindent$\bullet$ \textbf{Benchmark may not be comprehensive.} It is hard to thoroughly understand the buggy code and the patches in our selected open-source apps. Thus, our benchmark may not fully reflect the developers' usages of dangerous APIs or their practices of permission management. To address the threat, two authors worked together to understand the bugs/patches and built \benchn. We also made it publicly available. We hope that future researchers can help further improve \benchn, which may benefit the whole community.

\section{Related Work}
\label{sec:related}

We discuss three categories of related work in this section.

\noindent$\bullet$ \textbf{Mining Permission Specification.}
Building reliable Android permission specification has been extensively studied for many years~\cite{stowaway2011CCS,Copes2012ASE,au2012pscout,backes2016demystifying,karim2016mining,bogdanas2017dperm,aafer2018precise,dawoud2021bringing}.
Most existing studies employed either dynamic analyses (\eg, Stowaway \cite{stowaway2011CCS}) or static analyses (\eg, PScout \cite{au2012pscout}) to find API-permission mappings.
Among them, Stowaway, Copes \cite{Copes2012ASE}, PScout and Axplorer \cite{backes2016demystifying} were proposed before the adoption of the runtime permission model.
Arcade \cite{aafer2018precise} is a static approach proposed in 2018 for extracting permission specification.
It addressed the issue of imprecise mapping by handling path-sensitivity in the APIs.
Its mapping considers not just the permission protection, but other security attributes (\eg, API caller's UID/PID).
Dynamo \cite{dawoud2021bringing} is a dynamic approach that extracts permission mappings through API fuzzing.
In their paper, the authors stated that only 76.1\% of the common APIs reported by Dynamo and Arcade have matching security checks.
Since Dynamo builds permission mappings through dynamic testing, its result is supposed to be more precise.
However, these existing mappings are either outdated or incomplete: Arcade released the mappings up to API level 25, Dynamo only released the mappings in two API levels, 23 and 29.

Our mapping extraction was inspired by DPSpec \cite{bogdanas2017dperm}, 
which also extracts mappings from the annotations and Javadocs of the Android framework APIs.
However, their mappings were unavailable, and that was the reason we extracted the mappings by ourselves.

\noindent$\bullet$ \textbf{Runtime Permission Migration.}
Since the emergence of Android 6.0, many attempts have been made to migrate legacy apps to the new runtime permission model.
Most of them applied static analysis on an app's ICFG (or other variants) to decide proper program locations to insert calls to permission management APIs.
\textsc{ARPDroid} \cite{dilhara2018automated} inserts calls to the \textit{CHECK}, \textit{REQUEST}, and \textit{HANDLE} APIs into its identified incompatible permission-responsible callers.
However, it does not properly deal with the existing \textit{CHECK} and \textit{REQUEST} API call sites, and thus suffers from low precision and recall.
Gasparis \etal \cite{gasparis2018droid} pointed out that most developers considered migrating to the runtime permission model to be laborious.
To ease migration, they proposed \textsc{Droid M+} to migrate legacy apps with comprehensive runtime permission managements.

Previous methods were proposed under the assumption that the apps under processing target legacy platforms, thus they chose to actively insert calls to permission management APIs whenever a dangerous API call site lacks permission management.
In comparison, \tool only reports those dangerous contexts that are not dominated by permission or RV checks.
It helps developers find potential ARP bugs in their apps and avoid unexpected runtime behaviors, such as crashes.

\noindent$\bullet$ \textbf{ARP Bug Detection.}
Runtime permission migration has become less important in recent years as most apps have targeted new Android versions~\cite{WEB:android/target-sdk-version}. However, various ARP bugs are still lurking in apps and different techniques have been proposed to detect them.

\textsc{RevDroid}~\cite{fang2016revdroid} applies reachability analysis on the CG of an app to detect unexpected consequences after permission revocations.
Huang \etal~proposed a static method to detect stubborn permission request~\cite{huang2018detecting}, which repeatedly spawns a request dialog until the user grants that permission.
They also applied reachability analysis on the CG to detect such behaviors.
Unlike \tooln, these static methods do not fully consider asynchronous permission management, which is a common practice for app developers as revealed by our empirical study.
\textsc{RTPDroid} \cite{zhang2021rtpdroid} can detect the lack of permission checks before sensitive operations.
However, the work's main focus is on modeling and detecting implicitly malicious behaviors, thus falls into the security perspective.
We did not compare \tool with \textsc{RTPDroid} because the tool is not publicly available.

\textsc{PATDroid} \cite{sadeghi2017patdroid} applies hybrid analysis on an app to find crash-triggering test cases and permission combinations.
\textsc{Terminator} \cite{sadeghi2018temporal} also detects permission misuses via hybrid analysis, but it focuses on security issues.
\textsc{SetDroid} \cite{sun2021understanding} can detect system setting-related bugs by injecting setting-altering actions in the test event sequences. It is capable of finding bugs caused by permission revocation.
However, the performances of these test-driven approaches heavily rely on the underlying tests' coverage.
\tooln, as a static detector, models the permission usages and managements as four types of API invocation relations and comprehensively analyzes the existences of two types of ARP bugs.

To the best of our knowledge, \tool is the first tool that is able to detect ARP bugs caused by evolving permission specification.
Moreover, its modeling of the dangerous API calls and permission managements helps reduce false alarms and provide effective debugging information, as we have shown in the experiments.

\vspace{0.4em}
\section{Conclusion and Future Work}
\label{sec:conclusion}

In this paper, we studied the evolution of Android permission specification and real-world developers' permission management practices, via analyzing the Android framework source code and 13,352 top-ranked Android apps.
We found that both evolving dangerous APIs and asynchronous permission managements are widely used by real-world apps, which potentially bring ARP bugs but cannot be detected by existing tools.
To detect the two types of ARP bugs, we proposed \tooln, a new static analysis-based permission misuse analyzer.
We evaluated \tool with control experiments on a benchmark prepared by us, and an in-the-wild study on 214 open-source Android apps.
The results show that \tool can significantly outperform existing tools, and find real bugs in popular Android projects with useful debugging information.
In the future, we plan to extend \tool to support more types of runtime permission bugs, such as library-induced \cite{diamantaris2019reaper} or device-specific \cite{wei2019pivot} bugs.

\vspace{0.4em}
\begin{acks}
We would like to thank ICSE 2022 reviewers for their comments and suggestions, which helped improve this paper.
This work is partially supported by the National Natural Science Foundation of China (Grant Nos. 61932021, 61802164, 61902056), Guangdong Basic and Applied Basic Research Fund (Grant No. 2021A1515011562), Guangdong Provincial Key Laboratory (Grant No. 2020B121201001), Hong Kong GRF Project (No. 16211919), Hong Kong RGC Project (No. PolyU15223918), the Innovation and Technology Commission of Hong Kong (Innovation and Technology Fund MHP/055/19, PiH/255/21).
\end{acks}

\newpage

\balance
\bibliographystyle{ACM-Reference-Format}
\bibliography{web,pub}

\end{document}